# A Survey of Resource Management for Processing-in-Memory and Near-Memory Processing Architectures

**Kamil Khan, Sudeep Pasricha, and Ryan Gary Kim***

Department of Electrical and Computer Engineering, Colorado State University, Fort Collins, CO 80523, USA; Kamil.Khan@colostate.edu (K.K.); Sudeep.Pasricha@colostate.edu (S.P.); Ryan.G.Kim@colostate.edu (R.K.)

* Correspondence: Ryan.G.Kim@colostate.edu

**Abstract:** Due to amount of data involved in emerging deep learning and big data applications, operations related to data movement have quickly become the bottleneck. Data-centric computing (DCC), as enabled by processing-in-memory (PIM) and near-memory processing (NMP) paradigms, aims to accelerate these types of applications by moving the computation closer to the data. Over the past few years, researchers have proposed various memory architectures that enable DCC systems, such as logic layers in 3D stacked memories or charge sharing based bitwise operations in DRAM. However, application-specific memory access patterns, power and thermal concerns, memory technology limitations, and inconsistent performance gains complicate the offloading of computation in DCC systems. Therefore, designing intelligent resource management techniques for computation offloading is vital for leveraging the potential offered by this new paradigm. In this article, we survey the major trends in managing PIM and NMP-based DCC systems and provide a review of the landscape of resource management techniques employed by system designers for such systems. Additionally, we discuss the future challenges and opportunities in DCC management.

## 1. Introduction

For the past few decades, memory performance improvements have lagged compute performance improvements, creating an increasing mismatch between the time to transfer data and the time to perform computations on that data (the "memory wall"). The emergence of applications that focus on processing large amounts of data, such as deep machine learning and bioinformatics, have further exacerbated this problem. It is evident that the large latencies and energies involved with moving data to the processor will present an overwhelming bottleneck in future systems. To address this issue, researchers have proposed to reduce these costly data movements by introducing data-centric computing (DCC), where some of the computations are moved in proximity to the memory architecture.

Two major paradigms of DCC have emerged in recent years: processing-in-memory (PIM) and near-memory processing (NMP). In PIM architectures, characteristics of the memory are exploited and/or small circuits are added to memory cells to perform computations. For example, [1] takes advantage of DRAM's charge sharing property to perform bitwise PIM operations (e.g., AND and OR) by activating multiple rows simultaneously. These PIM operations allow computations to be done on memory where they are stored, thus eliminating most of the data movement. On the other hand, NMP architectures take advantage of existing compute substrates and integrate compute cores near the memory module. For example, modern 3D-stacked DRAM includes a logic layer where a compute core can be integrated beneath multiple DRAM layers within the same chip [2], [3].

Although the computation is done a little further away from memory than in PIM systems, NMP still significantly improves the latency, bandwidth, and energy consumption when compared to conventional computing architectures. For example, a commercial NMP chip called UPMEM implemented a custom core into conventional DRAM memory chips and achieved 25× better performance for genomic applications and 10× better energy consumption in an Intel x86 server compared to an Intel x86 server without UPMEM [4].

Both PIM and NMP systems have the potential to speed up application execution by reducing data movements. However, not all instructions can be simply offloaded onto the memory processor. Many PIM systems leverage memory characteristics to enable bulk bitwise operations, but other types of operations cannot be directly mapped onto the in-memory compute fabric. Even in NMP systems that utilize cores with full ISA support, performing computation in the 3D memory stack can create high power densities and thermal challenges. In addition, if some of the data has high locality and reuse, the main processor can exploit the traditional cache hierarchies and outperform PIM/NMP systems for instructions that operate on that data. In this case, the downside of moving data is offset by the higher performance of the main processor. All of these issues make it difficult to decide which computations should be offloaded and make use of these PIM/NMP systems.

In this article, we survey the landscape of different resource management techniques that decide which computations are offloaded onto the PIM/NMP systems. These management techniques broadly rely on code annotation (programmers select the sections of code to be offloaded), compiler optimization (compiler analysis of the code), and online heuristics (rule-based online decisions). Before providing a detailed discussion of resource management for PIM/NMP systems, Section 2 discusses prior surveys in PIM/NMP. Section 3 discusses various PIM/NMP design considerations. Section 4 discusses the optimization objectives and knobs, and different resource management techniques utilized to manage the variety of PIM/NMP systems proposed to date. Lastly, Section 5 concludes with a discussion of challenges and future directions.

## 2. Prior Surveys and Scope

Different aspects of DCC systems have been covered by other surveys. Siegl et al. [5] focus on the historical evolution of DCC systems from minimally changed DRAM chips to advanced 3D-stacked chips with multiple processing elements (PEs). The authors identify the prominent drivers for this change: first, memory, bandwidth, and power limitations in the age of growing big-data workloads make a strong case for utilizing DCC. Second, emerging memory technologies enable DCC, for example, 3D-stacking technology allows embedding PEs closer to memory chips than ever before. The authors identify several challenges with both PIM and NMP systems such as programmability, processing speed, upgradability, and commercial feasibility.

Singh et al. [6] focus on the classification of published work based on the type of memory, PEs, interoperability, and applications. They divide their discussion into RAM and storage-based memory. The two categories are further divided based on the type of PE used, such as fixed-function, reconfigurable, and fully programmable units. The survey identifies cache coherence, virtual memory support, unified programming model, and efficient data mapping as contemporary challenges. Similarly, Ghose et al. [7] classify published work based on the level of modifications introduced into the memory chip. The two categories discussed are 2D DRAM chips with minimal changes and 3D DRAM with one or more PEs on the logic layer.

Gui et al. [8] focus on DCC in the context of graph accelerators. Apart from a general discussion on graph accelerators, the survey deals with graph accelerators in memory and compares the benefits of such systems against traditional graph accelerators that use FPGAs or ASICs. It argues that memory-based accelerators can achieve higher bandwidth and low latency, but the performance of a graph accelerator also relies on other architectural choices and workloads.

In addition to DRAM, emerging memory technologies have been used to implement DCC architectures. Umesh et al. [9] survey the use of spintronic memory technology to design basic logic gates and arithmetic units for use in PIM operations. The literature is classified based on the type of operations performed, such as Boolean operations, addition, and multiplication. An overview of application-specific architectures is also included. Lastly, the survey highlights the higher latency and write energy with respect to SRAM and DRAM as the major challenges in large-scale adoption. Similarly, Mittal et al. [10] discuss resistive RAM (ReRAM) based implementations of neuromorphic and PIM logical and arithmetic units.

The surveys discussed above adopt different viewpoints in presenting prominent work in the DCC domain. But all of these surveys limit their discussion to the architecture and/or application of DCC systems and lack a discussion on the management techniques of such systems. With an increase in the architectural design complexity, we believe that the management techniques for DCC systems have become an important research area. Understandably, researchers have begun to focus on how to optimally manage DCC systems. In this survey, unlike prior surveys we therefore focus on exploring the landscape of management techniques that control the allocation of resources according to the optimization objectives and constraints in a DCC system. We hope that this survey provides the background and the spark needed for innovations in this critical component of DCC systems.

## 3. Data-Centric Computing Architectures

As many modern big-data applications require us to process massive datasets, large volumes of data are shared between the processor and memory subsystems. The data is too large to fit in on-chip cache hierarchies, therefore the resulting off-chip data movement between processors (CPUs, GPUs, accelerators) and main memory leads to long execution time stalls as main memory is relatively slow to service large influx of requests. The growing disparity in memory and processor performance is referred to as the memory wall. Data movement also consumes a significant amount of energy in the system, driving the system closer to its power wall. As data processing requirements continue to increase, the cost due to each wall will make conventional computing paradigms incapable of meeting application quality-of-service goals. To alleviate this bottleneck, PEs near or within memory have emerged as a means to perform computation and reduce data movement between traditional processors and main memory. Predominantly, prior work in this domain can be classified depending on how PEs are integrated with memory.

For DCC architectures, solutions can be divided into two main categories: 1) PIM systems, which perform computations using special circuitry inside the memory module or by taking advantage of particular aspects of the memory itself, e.g., simultaneous activation of multiple DRAM rows for logical operations [1], [11]–[25]; and 2) NMP systems, which perform computations on a PE placed close to the memory module, e.g., CPU or GPU cores placed on the logic layer of 3D-stacked memory [26]–[42]. For the purposes of this survey, we classify systems that use logic layers in 3D-stacked memories as NMP systems, as these logic layers are essentially computational cores that are near the memory stack (directly underneath it).

### 3.1. Processing-in-Memory (PIM) Designs

PIM solutions proposed to date typically leverage the high internal bandwidth of DRAM DIMMs to accelerate computation by modifying the architecture or operation of DRAM chips to implement computations within the memory. Beyond DRAM, researchers have also demonstrated similar PIM capabilities by leveraging the unique properties of emerging non-volatile memory (NVM) technologies such as resistive RAM (ReRAM), spintronic memory, and phase-change memory (PCM). The specific implementation approaches vary widely depending on the type of memory technology used, however, the modifications are generally minimal to preserve the original function of the memory unit and meet area constraints. For instance, some DRAM-based solutions minimally change the DRAM cell architecture while relying heavily on altering memory commands from the memory controller to enable functions such as copying a row of cells to another [16], logical operations such as AND, OR, and NOT [18], [22], [24], and arithmetic operations such as addition and multiplication [1], [23], [43]. Both DRAM and NVM-based architectures have demonstrated promising improvements for graph and database workloads [1], [18], [24], [44] by accelerating search and update operations directly where the data is stored. In the next sections, we will briefly discuss different PIM architectures that use DRAM and NVM.

*3.1.1. PIM Using DRAM*

The earliest PIM implementations [11], [14], [15] integrated logic within DRAM. The computational capability of these in-memory accelerators can range from simply copying a DRAM row [16] to performing bulk bitwise logical operations completely inside memory [1], [18], [23], [44]. For example, [1], [16], [23] use a technique called charge sharing to enable bulk AND and OR operations completely inside memory. Charge sharing is performed by the simultaneous activation of three rows called triple row activation (TRA). Two of the three rows hold the operands while the third row is initialized to zeros for a bulk AND operation or to ones for a bulk OR operation. Figure 1 shows an example TRA where cells A and B correspond to the operands (A is zero and B is one) while cell C is initialized to one to perform an OR operation ❶. The wordlines of all the three cells are raised simultaneously ❷, causing the three cells to share their charge on the bitline. Since two of the three cells are in a charged state, this results in an increase of voltage on the bitline. The sense amplifier ❸ then drives the bitline to $V_{DD}$ which fully charges all three cells completing the "A OR B" operation. In practice, this operation is done on many bitlines simultaneously to perform bulk AND or OR operations. In addition, with minimal changes to the design of the sense amplifier in the DRAM substrate, [1], [44] enable NOT operations in memory, which allows the design of more useful combinational logic for arithmetic operations like addition and multiplication. Besides bitwise operations, DRAM PIM has been shown to significantly improve neural network computation inside memory. For example, by performing operations commonly found in convolutional network networks like the multiply-and-accumulate operation in memory, DRAM PIM can achieve significant speed-up over conventional architectures [45].

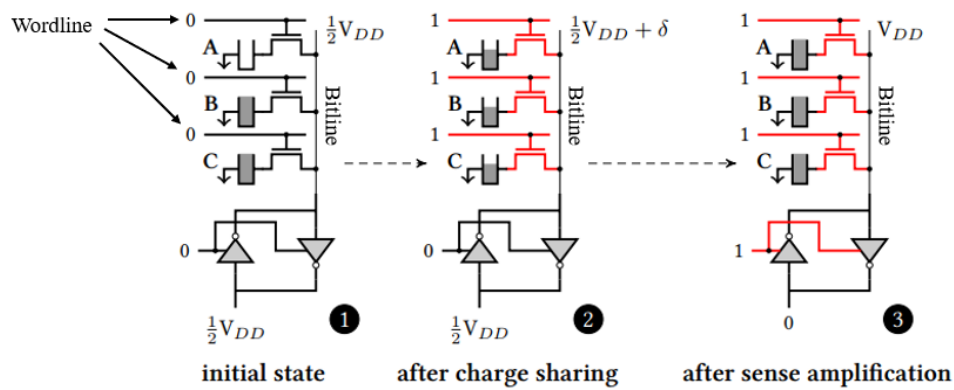

**Figure 1.** An example of an OR operation being performed on DRAM cells A and B using charge-sharing in triple-row activation [1].

In order to extract even more performance improvements, [12] places single instruction, multiple data (SIMD) PEs adjacent to the sense amplifiers at the cost of higher area and power per bit of memory. The inclusion of SIMD PEs inside the memory chip enables DRAM PIM to take advantage of the high internal DRAM bandwidth and minimizes the need for data to traverse power hungry off-chip links while enabling more complex computations than those afforded by TRA-based operations.

While PIM operations have been shown to outperform the host CPU (or GPU) execution, there are some technical challenges with their implementation. To enable these operations in DRAM, PIM modifications to DRAM chips are often made at the bank or sub-array level. If the source rows containing the operands do not share the same sensing circuitry, large amounts of data may need to be internally copied between banks, increasing latency and energy consumption, necessitating the use of partitioning and data-mapping techniques. In addition, DRAM PIM operations that use TRA are normally destructive, requiring several bandwidth consuming row copy operations if the data needs to be maintained *[1], [16]*. Alternative ways to accomplish computation inside DRAM chips include using a combination of multiplexers and 3T1C (three transistors, one capacitor) DRAM cells [46], [47]. While 3T1C cells allow non-destructive reads, these significantly increase the area overhead of the PIM system compared to the single transistor counterparts [1], [16], [44]. Moreover, DRAM is typically designed using a high threshold voltage process which slows down PIM logic operations when PIM PEs are fabricated on the same chip [48]. These factors have resulted in relatively simple DRAM PIM designs that prevent entire applications from running entirely in memory.

*3.1.2. PIM Using NVM*

Charge-storage-based memory (DRAM) has been encountering challenges in efficiently storing and accurately reading data as transistor sizes shrink and operating frequencies increase. DRAM performance and energy consumption has not scaled proportionally with transistor sizes like it did for processors. Moreover, the charge storage nature of DRAM cells requires that data be periodically refreshed. Such refresh operations lead to higher memory access latency and energy consumption. Kim et al. [49] have also demonstrated that contemporary DRAM designs suffer from problems such as susceptibility to RowHammer attacks [50] which exploit the limitations of charge-based memory to induce bit-flip errors. Solutions to overcome such attacks can further increase execution time and reduce the energy efficiency of DRAM PIM designs.

On the other hand, alternative NVM memory technologies such as phase-change memory (PCM) [51]–[54], ReRAM [22], [24], [55]–[59], and spintronic RAM [60]–[66] show promise. NVM eliminates the reliance on charge memory and represents the data as cell resistance values instead. When NVM cells are read, cell resistances are compared to reference values to ascertain the value of the stored bit, i.e., if the measured resistance is within a preset range of low resistance values ($R_{low}$), the cell value read is a logical "1", whereas a cell resistance value within the range of high resistance values ($R_{high}$) is read as a logical "0". Writing data involves driving an electric current through the memory cell to change its physical properties, i.e., the material phase of the crystal in PCM, the magnetic polarity in spintronic RAM, and the atomic structure in memristors and ReRAM. The data to be stored in all cases is embedded in the resultant resistance of the memory cell, which persists even when the power supply shuts off. Thus, unlike DRAM, such memories store data in a non-volatile manner. In addition to being non-volatile, the unique properties of these memory technologies provide higher density, lower read power, and lower read latency compared to DRAM.

PIM using NVM, while based on a fundamentally different memory technology, can resemble its DRAM counterpart. For example, Li et al. [24] activate multiple rows in a resistive memory sub-array to enable bitwise logical operations. By activating multiple rows, the sense amplifiers measure the bitwise parallel resistance of the two rows on the bitlines and compare the resistance with a preset reference resistance to determine the output. Figure 2(a) shows how any two cells, R1 and R2, on the same bitline connect in parallel to produce a total resistance R1||R2 where || refers to parallel resistance of two cells. Figure 2(b) shows how the parallel resistance formed by input resistances ($R_{low}$ or $R_{high}$) measured at the sense amplifier, together with a reference value (RrefOR/RrefAND) can be used to perform the logical AND and OR operation in memory [24].

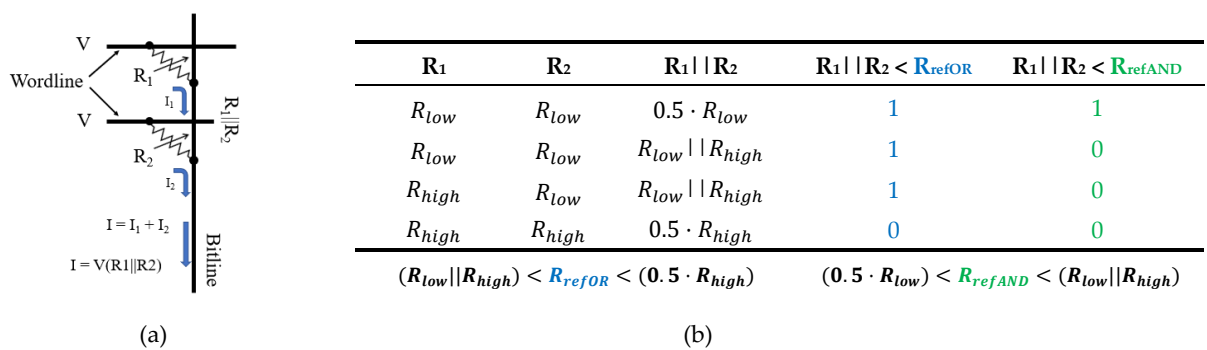

**Figure 2.** Example of AND and OR operations in memory that use resistance-based memory cells. (a) Activating two resistors R1 and R2 on the same bitline in a memristive crossbar array (MCA) results in an effective resistance of R1||R2 on the bitline (Figure adapted from [56]). This effective resistance can be compared against preset reference resistance values (RrefOR, RrefAND) to ascertain the result of the Boolean operation, (b) shows the truth table [24].

Aside from the inherent benefits of NVM technology, e.g., energy efficient reads, persistence, and latency, NVM-based PIM allows AND/OR operations on multiple rows as opposed to just two in DRAM. This is because as long as each memory cell has a large range of resistances, resistive memory can have multiple levels of resultant resistances that can represent multiple states. Another technological advantage for NVM PIM is that unlike DRAM, reads are not destructive. This eliminates the need to copy data to special rows before operating on them, as done in DRAM based PIM solutions.

Besides conventional logical operations, NVM PIM has also proven particularly useful for accelerating neural network (NN) computations [9], [10]. Cell conductance in the rows of memristive crossbar arrays (MCA) can be used to represent synaptic weights and the input feature values can be represented by wordline voltages. Then, the current flowing through each bitline will be related to the dot product of input values and weights in a column. As this input-weight dot product is a key operation for NNs, PIM architects have achieved significant acceleration for NN computation [19] by extending traditional MCA to include in-memory multiply operations. Other applications that have been shown to benefit from NVM PIM include data search operations [67], [68] and graph processing [57].

A significant drawback of using NVM for PIM is area overhead. The analog operation of NVM requires the use of DAC (digital-to-analog convertor) and ADC (analog-to-digital converter) interfaces. This results in considerable area overhead. For example, DAC and ADC converters in the implementation of a 4-layer convolutional neural network can take up to 98% of the total area and power of the PIM system [69].

In summary, PIM designs typically allow for very high internal data bandwidth, greater energy efficiency, and low area overhead by utilizing integral memory functions and components. However, to keep memory functions intact, these designs can only afford minimal modifications to the memory system which typically inhibits the programmability and computational capability of the PEs.

*3.2. Near-Memory Processing (NMP) Designs*

NMP designs utilize more traditional PEs that are placed near the memory. However, as the computations are not done directly in memory arrays, PEs in NMP do not enjoy the same degree of high internal memory bandwidth available to PIM designs. NMP designs commonly adopt a 3D stacked structure such as that provided by hybrid memory cube (HMC) [2] or high bandwidth memory (HBM) [3]. Such designs have coarser-grained offloading workloads than PIM but can use the logic layer in 3D stacks to directly perform more complex computations. In the following sections we discuss the different PE and memory types that have been explored for NMP systems.

*3.2.1 PE Types*

Many different PE types have been proposed for NMP designs. The selection of PEs has a significant impact on the throughput, power, area, and types of computations that the PE can perform. For example, Figure 3 compares the throughput of different PE types in an HMC-based NMP system executing a graph workload (PageRank scatter kernel). The throughput is normalized to the maximum memory bandwidth which is indicated by the solid line. This indicates that while multi-threaded (MT) and SIMD cores allow flexibility, they cannot take advantage of all the available bandwidth. On the other hand, ASICs easily saturates the memory bandwidth and become memory bound. In this particular situation, reconfigurable logic like FPGA and CGRA may be better to balance performance and flexibility. However, in the end, this tradeoff point depends on the application, memory architecture, and available PEs. Therefore, the choice of PE is an important design consideration in NMP systems. We highlight the three types of PEs below.

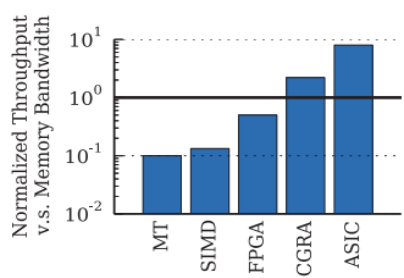

**Figure 3.** Normalized throughput with respect to maximum memory bandwidth for different PEs executing the PageRank scatter kernel: multi-threaded (MT) programmable core, SIMD programmable core, FPGAs, CGRAs, and fixed-function ASICs [41].

**Fixed function accelerators:** Fixed function accelerator PEs are ASICs designed with support for a reduced set of operations to speed up a particular application or task, for example, graph processing [8], [32], [57]. These accelerators are highly specialized in their execution, which allows them to achieve much greater throughput than general purpose processors for the same area and power budget. However, it is costly and challenging to customize an accelerator for a target workload. As these accelerators only execute a limited set of instructions, they cannot be easily retargeted to new (or new versions of existing) workloads.

**Programmable logic:** Programmable logic PEs can include general-purpose processor cores such as CPUs [31], [40], [70]–[72], GPUs [26], [27], [38], [73], [74], and APUs [75] that can execute complex workloads. These cores are usually trimmed down (fewer computation units, less complex cache hierarchies without L2/L3 caches, or lower operating frequencies) from their conventional counterparts due to power, area, and thermal constraints. The main advantage of using these general-purpose computing cores is flexibility and programmability. As opposed to fixed function accelerators, NMP systems with programmable units allow any operation normally executed on the host processor to potentially be performed near memory. However, this approach is hampered by several issues such as difficulty in maintaining cache coherence [31], implementing virtual address translation [76], and meeting power, thermal and area constraints [29], [73], [77]–[79].

**Reconfigurable logic:** Reconfigurable logic PEs include CGRAs and FPGAs that can be used to dynamically change the NMP computational unit hardware [21], [25], [41], [80]. In NMP systems, such PEs act as a compromise between the efficiency of simple fixed-function memory accelerators and the flexibility of software-programmable complex general-purpose cores. However, the reconfiguration of hardware logic results in runtime overhead and additional system complexity.

*3.2.2 Memory Types*

The type of memory to use is an important decision for DCC system designers working with NMP systems. Different memory technologies have unique advantages that can be leveraged for particular operations. For example, write-heavy operations are more suited to DRAM than resistive memories due to the larger write latency of resistive memories. This fact and DRAM's widespread adoption has led to the creation of a commercial DRAM NMP solution UPMEM [4]. UPMEM replaces standard DIMMs, allowing it to be easily adopted in datacenters. A 3D-stacked DRAM architecture like the HMC [81] can accelerate atomic operations of the form read-modify-write (RMW) due to the memory level parallelism supported by the multi-vault, multi-layer structure shown in Figure 4. In fact, NMP designs prefer 3D-DRAM over 2D-DRAM due to several reasons. First, 3D-DRAM solutions such as HMC 2.0 [81] have native support for executing simple atomic instructions. Second, 3D-DRAM allows for easy integration of a logic die that provides greater area for placing more complex PEs near memory than 2D counterparts. Third, PEs are connected using high bandwidth through-silicon vias (TSVs) to memory arrays as opposed to off-chip links, providing much higher memory bandwidth to the PE. Fourth, the partitioning of memory arrays into vaults enables superior memory-level parallelism.

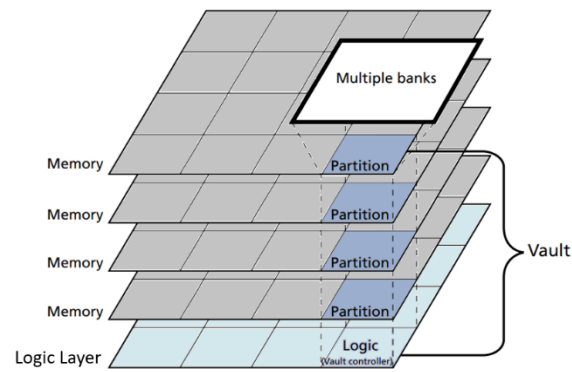

**Figure 4.** The architecture of Hybrid Memory Cube (HMC) [81].

Resistive memory in its memristive crossbar array (MCA) configuration can significantly accelerate matrix multiply operations commonly found in neural network, graph, and image processing workloads [19], [82]–[89]. Figure 5(a) shows a mathematical abstraction of one such operation for a single-layer perceptron implementing binary classification, where $x$ are the inputs to the perceptron, $w$ are the weights to the perceptron, $sgn$ is the signum function, and $y$ is the output of the perceptron [90]. The MCA is formed by non-volatile resistive memory cells placed at each crosspoint of a crossbar array. Figure 5(b) shows how the perceptron can be mapped onto an MCA. The synaptic weights of the perceptron, $w$, are represented by physical cell conductance values, $G$. Specifically, each weight value is represented by a pair of cell conductances, i.e., $w_i \propto G_i \equiv G_i^+ - G_i^-$ to represent both positive and negative weight values. The inputs, $x$, are represented by proportional input voltage values, $V$, applied to the crossbar columns. For the device shown, $V = \pm 0.2$ V. In this configuration, the weighted sum of the inputs and synaptic weights ($y = \sum_{i=0}^{9} w_i x_i$) can be obtained by reading the current at the each of the two crossbar rows ($I^+ = \sum_{i=0}^{9} G_i^+ V_i$ and $I^- = \sum_{i=0}^{9} G_i^- V_i$). Finally, the sign of the difference in the current flowing through the two rows ($sgn[\sum_{i=0}^{9} G_i^+ V_i - \sum_{i=0}^{9} G_i^- V_i]$) can be determined which is equivalent to the original output $y$. In this manner, MCA is able to perform the matrix multiplication operation in just one step.

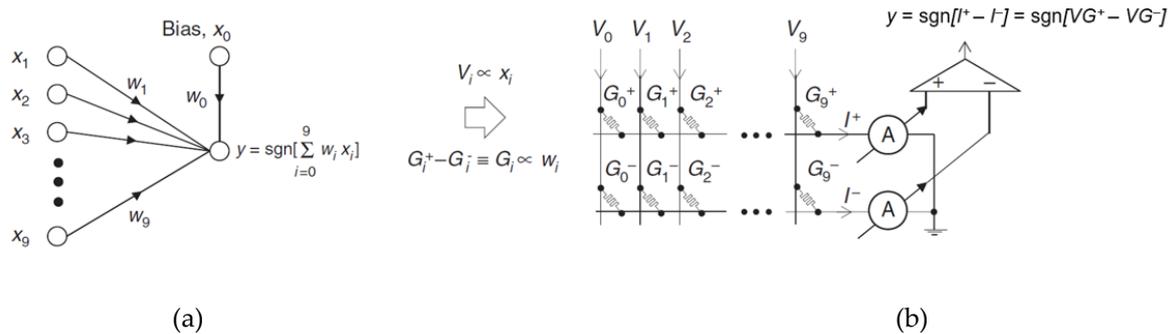

(a)  (b)

**Figure 5.** (a) A mathematical abstraction of a single-layer perceptron used for binary classification. (b) A memristive crossbar array implementing the weighted sum between inputs $x_i$ (mapped to wordline voltages $V_i$) and weights $w_i$ (mapped to cell conductance $G_i$) [90].

While the choice of a PE and memory type are important factors in determining the advantages of an NMP system over a PIM system, some features are shared by most NMP designs. Due to the external placement of the PE, NMP can afford a larger area and power budget compared to PIM designs. However, NMP also has lower maximum data bandwidth available to the PE compared to PIM systems.

*3.3. Data Offloading Granularity*

For both PIM and NMP systems, it is important to determine what computation will be sent (offloaded) to the memory PE. Offloading can be performed at different granularities, e.g., instructions (including small groups of instructions) [1], [13], [16], [19], [24], [25], [28], [32], [37], [39], [40], [42], [57], [91], [92], threads [71], CUDA blocks/warps [27], [29], kernels [26], and applications [38], [41], [73], [74]. Instruction-level offloading is often used with a fixed-function accelerator and PIM systems [1], [13], [16], [19], [24], [25], [28], [29], [32], [37], [39], [42], [57], [92]. For example, [42] offloads atomic instructions at instruction-level granularity to a fixed-function near-memory graph accelerator. Coarser offloading granularity, such as kernel and full application, requires more complex memory PEs that can fetch and execute instructions, and maintain a program counter. Consequently, coarse-grained offloading is often used in conjunction with programmable memory PEs such as CPUs [71], [93] and GPUs [26], [27], [38], [73], [74].

## 4. Resource Management of Data-Centric Computing Systems

Although it would be tempting to offload all instructions to the PIM or NMP system and eliminate most data movements to/from the host processor, there are significant barriers to doing so. First, memory chips are usually resource constrained and cannot be used to perform unrestricted computations without running into power/thermal issues. For example, 3D-DRAM NMP systems often place memory processors in the 3D memory stack that can present significant thermal problems if not properly managed. Figure 6 [29] demonstrates how the peak temperature of a 3D-stacked DRAM with a low-power GPU on the logic layer changes with the frequency of offloaded operations (PIM rate). Consistent operation at 2 op/ns will result in the DRAM module generating a thermal warning at 85 °C. Depending on the DRAM module, this can lead to a complete shutdown, or degraded performance under high refresh rates. If the PIM rate is unmanaged, the DRAM cannot guarantee reliable operation due to the high amount of charge leakage under high temperature.

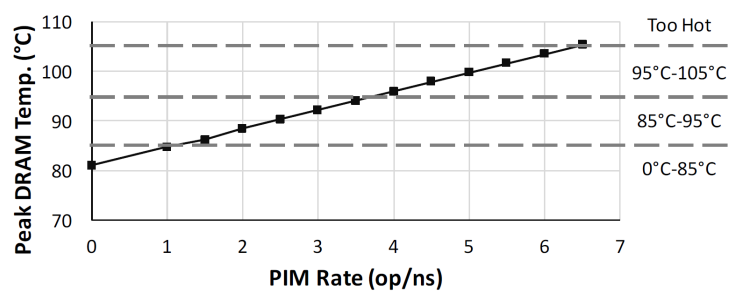

**Figure 6.** Peak DRAM temperatures at different PIM offloading frequencies (PIM Rate) in a 3D-DRAM NMP system [29].

Secondly, memory processors are not guaranteed to improve power and performance for all workloads. For example, workloads that exhibit spatial or temporal locality in their memory accesses are known to perform worse in PIM/NMP systems [26], [28]. When executed on a host processor, these workloads can avoid excessive DRAM accesses and utilize the faster, more energy efficient on-chip cache. Only the least cache-friendly portions of an application should be offloaded to the memory processors while the cache-friendly portions are run on the host processor. Figure 7 [28] compares the speedup achieved for a "Host-Only" (entire execution on the host), "PIM-Only" (entire execution on the memory processor), and "Locality-Aware" (offloads low locality instructions) offloading approaches to an NMP system for different input sizes. It should be noted that in [28], PIM refers to an HMC-based architecture we classify as NMP. For small input sizes, all except one application suffers from severe degradation with "PIM-Only". This is because the working set is small enough to fit in the host processor's cache, thus achieving higher performance than memory-side execution. As input sizes grow from small to medium to large, the performance benefit of offloading is realized. Overall, the locality aware management policy proves to be the best of the three approaches and demonstrates the importance of correctly deciding which instructions to offload based on locality. Another useful metric to consider when making offloading decisions is the expected memory bandwidth-saving from the offloading [38], [42], [94].

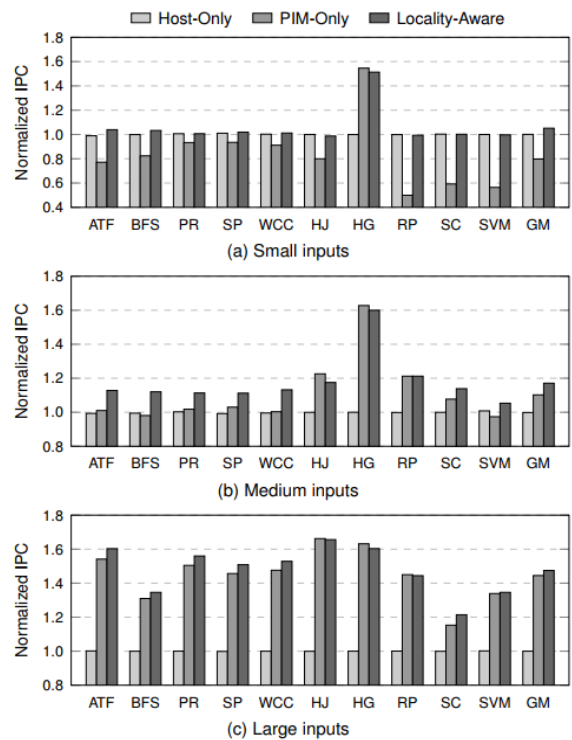

**Figure 7.** The speedup achieved by a Host-Only, all offloaded (PIM-Only), and low-locality offload (Locality-Aware) approaches in a DRAM NMP system for different application input sizes [28].

Lastly, these DCC systems can include different types of PE and/or multiple PEs with different locations within the memory. This can make it challenging to determine which PE to select when specific instructions need to be offloaded. For example, NMP systems may have multiple PEs placed in different locations in the memory hierarchy [94] or in different vaults in 3D-based memory [26]. This will result in PEs with different memory latency and energy depending on the location of the data accessed by the computation. In these cases, it is vital that the correct instructions are offloaded to the most suitable candidate PE.

To address and mitigate the above issues in DCC systems, offloading management policies are necessary to analyze and identify instructions to offload. Like every management policy, the management of DCC systems can be divided into three main ideas: 1) defining the optimization objectives, 2) identifying optimization knobs or the parts of the system that the policy can alter to achieve its goals, and 3) defining a management policy to make the proper decisions. For the optimization objectives, performance, energy, and thermals have been targeted for PIM and NMP systems. The most common optimization knobs in DCCs include selecting offloading workloads for memory, selecting the most suitable PE in/near memory, or the timing of executing selected offloads. To implement the policy, management techniques have employed code annotation [1], [13], [16], [19], [24], [25], [28], [31], [32], [37], [40], [57], [91], [95], compiler-based code analysis [27], [39], [40], [70], [92], [96], and online heuristics [27]–[29], [38], [71], [72], [74]. Table 1 classifies prominent works based on these attributes. We further discuss the optimization objectives, optimization knobs, and management techniques in Sections 4.1, 4.2, and 4.3, respectively.

*4.1 Optimization Objectives*

An optimization objective is pivotal to the definition of a resource management policy. Although the direct goal of PIM/NMP systems is to reduce data-movement between the host and memory, the optimization objectives are better expressed as performance improvement [1], [13], [16], [19], [22],

[24], [25], [31], [32], [37]–[41], [57], [59], [71]–[73], [75], [92], [96], energy efficiency [38], [73], [91], and thermal feasibility [29] of the system.

**Table 1.** Classification of resource management techniques for DCC systems (E: Energy, P: Performance, Pow: Power, T: Temperature)

| Management Method | Objectives | Architecture | Offload Granularity | Work (Year) |
|---|---|---|---|---|
| Code Annotation | E | NMP | Instruction | [91] (2014) |
|  | P | NMP | Group of instructions | [31] (2017) |
|  |  | PIM | Instruction | [13] (1995); [24] (2016); [25] (2017) |
|  | P/E | NMP | Instruction | [28] (2015); [32], [37] (2017); [40], [57] (2018) |
|  |  | NMP | Group of instructions | [75] (2014); [95] (2016) |
|  |  | PIM | Instruction | [16] (2013); [19] (2016); [1] (2017) |
| Compiler | P | NMP | Instruction | [42] (2017); [92] (2019) |
|  |  | NMP | Group of instructions | [96] (2015); [70] (2017) |
|  | P/E | NMP | Instruction | [40] (2018); [39] (2019) |
|  |  | NMP | Thread | [27] (2016) |
| Online - Heuristic | P | NMP | Thread | [71] (2018) |
|  | P/E | NMP | Instruction | [28] (2015) |
|  |  | NMP | Group of instructions | [72] (2020) |
|  |  | NMP | Thread | [27] (2016); [74] (2019) |
|  |  | NMP | Application | [26] (2016) |
|  | Pow/P/E | NMP | Application | [38] (2020) |
|  | T | NMP | Instruction | [29] (2018) |

*4.1.1 Performance*

DCC architectures provide excellent opportunities for improving performance by saving memory bandwidth, avoiding cache pollution, exploiting memory level parallelism, and using special-purpose accelerators. Management policies are designed to take advantage of these according to the specific DCC design and workload. For example, locality-aware instruction offloading [26]–[28], [38], [42], [74] involves an analysis of the cache behavior of a set of instructions in an application. This analysis is based on spatial locality (memory locations are more likely to be accessed if neighboring memory locations have been recently accessed) and temporal locality (memory locations are more likely to be accessed if recently accessed). If a section of code's memory access pattern is found to exhibit little spatial and temporal locality, then these instructions are considered a good candidate for offloading to a memory PE. The specific method to assess locality varies across different implementations. Commonly applied methods include cache profiling [27], [28], [42], code analysis [26], [39], [42], [70], [96], hardware cache-hit counters [28], and heuristic or machine learning techniques [27]–[29], [38], [71], [72], [74].

Another source of superior performance is the use of fixed-function accelerators. While lacking programmability, these special-purpose accelerators can perform specific operations many times faster than general-purpose host processors. Examples include graph accelerators [8], [32]–[36], [57], stencil computation units [39], texture filtering units [37], vector processing units [92], and neuromorphic accelerators [25]. When such fixed-function accelerators are used, it is important that the policy offload only instructions that the fixed-function accelerator is able to execute.

Finally, in the case of multiple PEs, data placement and workload scheduling become important to realizing superior performance. For example, the compiler-based offloading policy outlined in [92] selects the PE which minimizes the resulting inter-PE communication resulting from accessing data from different memory vaults during execution. Similarly, [97] uses compiler and profiling techniques to map operations to PEs that minimize data movement between memory vaults in a 3D-stacked HMC.

*4.1.2 Energy Efficiency*

DCC architectures have demonstrated tremendous potential for energy efficiency due to a reduction in expensive off-chip data movement, the use of energy-efficient PEs, or by using NVM instead of leakage-prone DRAM. Off-chip data movement between a host CPU and main memory is found to consume up to 63% of the total energy spent in some consumer devices [40]. The excessive energy consumption can be significantly reduced by using locality-aware offloading, as discussed in the previous section. A related metric used by some policy designs is bandwidth saving. This is calculated by estimating the change in the total number of packets that traverse the off-chip link due to an offloading decision about a set of instructions. For example, in [42], the cache-hit ratio and frequency of occurrence of an instruction is considered along with the memory bandwidth usage to decide if an instruction should be offloaded.

Another source of energy-efficiency is fixed-function accelerators. Since these accelerators are designed to execute a fixed set of operations, they do not require a program counter, load-store units, excessive number of registers and ALUs, which significantly reduces both static and dynamic energy consumption. Using NVM can eliminate the cell refresh energy consumption in DRAM, due to the inherent non-volatility in NVMs. In addition, NVMs that use an MCA architecture can better facilitate the matrix multiplication operation, requiring only one step to calculate the product of two matrices. In contrast, DRAM requires multiple loads/stores and computation steps to perform the same operation. This advantage is exploited by management policies which map matrix vector multiplication to ReRAM crossbar arrays [19], [82]–[89].

*4.1.3 Power and Thermal Efficiency*

DCC systems require power to be considered as both a constraint and an optimization objective, motivating a range of different management techniques from researchers. The memory module in particular is susceptible to adverse effects due to excessive power usage. It can lead to high temperatures, which may result in reduced performance, unreliable operation, and even thermal shutdown. Therefore, controlling the execution of offloaded workloads becomes important. Various ways in which power efficiency/feasibility is achieved using policy design is by using low-power fixed-function accelerators, smart scheduling techniques for multiple PEs, and thermal-aware offloading. For example, when atomic operations are offloaded to HMC, it reduces the energy per instruction and memory stalls compared to host execution [28], [29], [32]. Data mapping and partitioning in irregular workloads like graph processing reduces peak power and temperature when multiple PEs are used [36]. A more reactive strategy is to use the memory module's temperature as a feedback to the offloading management policy [29] to prevent offloading in the presence of high temperatures.

*4.2 Optimization Knobs*

There are several knobs available to the resource management policy to achieve the goals defined in the optimization objective. Typically, the management policy identifies which operations to offload (we call the operations that end up being offloaded as <u>offloading workloads</u>) from a larger set of offloading candidates, i.e., all workloads that the policy determined can *potentially* run on a memory PE. If multiple PEs are available to execute offloading workloads, a selection must be made between them. Finally, decisions may involve timing offloads, i.e., when to offload workloads.

*4.2.1 Identification of Offloading Workloads*

The identification of offloading workloads concerns the selection of single instructions or groups of instructions which are suitable for execution on a memory-side processor. Depending on the computation capability of the available processor, the identification of offloading workloads can be accomplished in many ways. For example, an atomic unit for executing atomic instructions is used as the PE in [32], then the offloading candidates are all atomic instructions. The identification of offloading workloads is simply identifying which offloaded atomic instructions best optimizes the objectives. However, the process is not always as simple, especially when the memory-side processors offer a range of computation options, each resulting in different performance and energy costs. This can happen because of variation in state variables such as available memory bandwidth, locality of memory accesses, resource contention, and the relative capabilities of memory and host processors. For example, [26] uses a regression-based predictor for identifying GPU kernels to execute on a near-memory GPU unit. Their model is trained on kernel level analysis of memory intensity, kernel parallelism, and shared memory usage. Therefore, while single instructions for fixed-function accelerators can be easily identified as offloading workloads by the programmer or compiler, identifying offloading workloads for more complex processors in a highly dynamic system requires online techniques like heuristic or machine-learning based methods.

*4.2.2 Selection of Memory PE*

A common NMP configuration places a separate PE in each memory vault of a 3D-stacked HMC module [34], [92], [97]. This can lead to different PEs encountering different memory latencies when accessing operand data residing in different memory vaults. In addition, the bandwidth available to intra-vault communication is much higher (360 GB/s) than that available to inter-vault communication (120 GB/s) [34]. Ideally, scheduling decisions should aim to utilize the higher intra-vault bandwidth. An even greater scheduling challenge is posed by the use of multiple HMC modules connected by a network [34], [35]. The available bandwidth for communication between different HMC modules is even lower at 6 GB/s. Therefore, offloading decisions must consider the placement of data and the selection of PEs together to maximize performance [34], [35], [98]. Similarly, a reconfigurable unit may require reconfiguration before execution can begin on the offloaded workload [41].

Finally, the selection of the PE can be motivated by load-balancing, especially in graph processing where workloads are highly unstructured, e.g., when multiple HMC modules are used for graph processing acceleration. In this case, the graph is distributed across the different memory modules. Since graphs are typically irregular, this can cause irregular communication between memory modules and imbalanced load in the memory PEs [34], [35]. If load distribution is not balanced by PE selection, it can lead to long waiting times for under-utilized cores. These issues influence offloading management, often requiring code annotation/analysis and runtime monitoring.

*4.2.3 Timing of Offloads*

While identifying offloading workloads and selecting the PE for execution are vital first steps, the actual execution is subject to the availability of adequate resources and power/thermal budget. Offloading to memory uses off-chip communication from the host processor to the memory as an essential step to convey the workload for execution. If available off-chip bandwidth is currently limited, the offloading can be halted, and the host processor can be used to perform the operations until sufficient bandwidth is available again [26], [27], [38], [42]. Alternatively, computation can be "batched" to avoid frequent context switching [34]. Moreover, using the memory-side PE for computation incurs power and thermal cost. If it is predicted that the memory and processor system will violate power or thermal constraints while the offloading target is executed, offloading may be deferred or reduced to avoid degrading memory performance or violating thermal constraints [29]. Hence, in a resource constrained PIM/NMP system, naively offloading all offloading candidates can lead to degraded performance as well as power/thermal violations. Therefore, a management policy should be able to control not just what is being offloaded but also when and where it will be offloaded.

*4.3 Management Techniques*

In order to properly manage what is offloaded onto PIM or NMP systems, where it is offloaded (in the case of multiple memory PEs), and when it is offloaded, prior work utilizes one of three different strategies: (1) **code annotation**: techniques that rely on the programmers to select and determine the appropriate sections of code to offload; (2) **compiler optimization**: techniques that attempt to automatically identify what to offload during compile-time; and (3) **online heuristics**: techniques that use a set of rules to determine what to offload during run-time. We discuss the existing works in each of the three categories in the following sections.

*4.3.1 Code Annotation Approaches*

Code annotation is a simple and cost-effective way to identify offloading workloads with the programmer's help. For these techniques to work, the programmer must manually identify offloading workloads based on their expert knowledge of the underlying execution model while ensuring the efficient use of the memory module's processing resources. Although this approach allows for greater consistency and control over the execution of memory workloads, it burdens the programmer and relies on the programmer's ability to annotate the correct instructions. This policy has been demonstrated to work well with fixed-function units [1], [16], [24], [28], [32], [37] since the instructions that can be offloaded are easy to identify and the decision to offload is often guaranteed to improve the overall power and performance without violating thermal constraints. For example, [16] introduces two simple instructions for copying and zeroing a row in memory. The in-memory logic implementation of these operations is shown to be both faster and more energy-efficient, motivating the decision to offload all such operations. Code annotation has been successfully used in both PIM [1], [25], [98] and fixed-function NMP [28], [32], [40], [57].

Although the most popular use of code annotation is found in PIM or fixed-function NMP, code annotation's simplicity also makes it appealing to more general architectures. This is especially true when the policy focuses more on NMP-specific optimizations such as cache-coherence, address translation, or logic reconfiguration than the system-wide optimizations outlined in Section 4.1. For example, code annotation is used in architectures with fully programmable APUs [75], CPUs [31], and CGRAs [95]. In [31], an in-order CPU core is used as a PE in 3D-stacked memory. The task of offloading is facilitated by the programmer selecting portions of code by using macros *PIM_begin* and *PIM_end*, where the main goal of the policy is to reduce coherence traffic between the host and memory PE. For more general-purpose PEs and a greater number of PEs, the offloading workload becomes harder to identify with human expertise.

One major downside of code annotation approaches is that the offline nature of code annotations prevents it from adapting to the changes in the system's state, i.e., the availability of bandwidth, the cache locality of target instructions, and power and temperature constraints. To address this, code annotation has been used as a part of other management techniques like compiler-based methods [40] and online heuristics [28] that incorporate some online elements. Other variations of the technique include an extension to the C++ library for identifying offloadable operations [24], [98], extensions to the host ISA for uniformity across the system [16], [19], [57], [91], or a new ISA entirely [25]. Extending the software interface in this manner is particularly useful when the source code is not immediately executable on memory PEs or requires the programmer to repeatedly modify large sections of code. For example, a read-modify-write operation, expressed with multiple host instructions can be condensed into a single HMC 2.0 atomic instruction by introducing a new instruction to the host processor's ISA. Similarly, library extensions provide compact functions for ease of use and readability.

*Case Study 1: Ambit: In-Memory Accelerator for Bulk Bitwise Operations Using Commodity DRAM Technology*

Bulk bitwise operations like AND and OR are increasingly common in modern applications like database processing but require high memory bandwidth on conventional architectures. Ambit [1] uses a 2D-DRAM-based PIM architecture for accelerating these operations using the high internal memory bandwidth in the DRAM chip. To enable AND and OR operations in DRAM, Ambit proposes triple row activation (TRA). TRA implements bulk AND and OR operations by activating three DRAM rows simultaneously and taking advantage of DRAM's charge sharing nature (discussed in detail in Section 3.1.1). As these operations destroy the values in the three rows, data must be copied to the designated TRA rows prior to the Boolean operation if the original data is to be maintained.

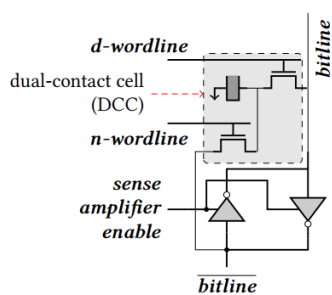

**Figure 8.** Ambit's implementation of the NOT operation by adding a dual-contact DRAM cell to the sense amplifier. The grey shaded region highlights the dual-contact cell [1].

To allow more flexible execution, Ambit also implements the NOT operation through a modification to the sense amplifiers shown in Figure 8. The NOT operation uses a dual-contact DRAM cell added to the sense amplifier to store the negated value of the cell in a capacitor and store it on the bitline when required. To perform the NOT operation, the value of a cell is read into the sense amplifier by activating the desired row and the bitline. Next, the n-wordline, which connects the negated cell value from the sense amplifier to the dual-contact cell capacitor is activated, which allows the capacitor to store the negated value. Finally, activating the d-wordline drives the bitline to the value stored in the dual-contact cell capacitor. AMBIT then uses techniques adapted from [16] to transfer the result to an operand row for use in computation. By using the memory components with minimal modifications, Ambit adds only 1% to the memory chip area, and allows easy integration using the traditional DRAM interface.

```
bbop dst, src1, [src2], size
```

**Figure 9.** Ambit [1] extends the ISA to include a bulk bitwise operation (bbop) instruction for computation which will be offloaded to PIM. The format of the instruction is shown.

To execute the Boolean operations, Ambit adds special bulk bitwise operation (bbop) instructions of the form shown in Figure 9 to the host CPU's ISA. These bbop instructions specify the type of operation (bbop), the two operand rows (src1 and src2), a destination row (dst), and the length of operation in bytes (size). Since the Ambit operations operate on entire DRAM rows, the size of the bbop operation must be a multiple of the DRAM row size. Due to Ambit's small set of operations, it is relatively easy to identify the specific code segments that can benefit from these operations. However, the authors expect that the implementers of Ambit would provide API/driver support that allows the programmer to perform code annotation to specify bitvectors that are likely to be involved in bitwise operations and have those bitvectors be placed in subarrays such that corresponding portions of each bitvector are in the same subarray. This is key to enabling TRA-based operations.

Applications that use bulk bitwise operations show significant improvement using Ambit. Ambit accelerates a join and scan application by 6× on average for different data sizes compared to a baseline CPU execution. Ambit accelerates the Bitweaving (database column scan operations) application by 7× on average compared to the baseline, with the largest improvement witnessed for large working sets. A third applications uses bitvectors, a technique used to store set data using bits to represent the presence of an element. Ambit outperforms the baseline for the application by 3× on average, with performance gains increasing with set sizes. Ambit also demonstrates better energy-efficiency compared to traditional CPU execution of applications using bulk bitwise operations. While raising additional wordlines consumes extra energy, the reduction of off-chip data movement more than makes up for it, reducing energy consumption by 25.1× to 59.5×.

While the architecture of Ambit shows significant promise, its management policy is not without issues. The benefits would rely heavily on the programmer's expertise as the programmer would still need to identify and specify which bitvectors are likely to be involved in these bitwise operations. Fortunately, this technique avoids runtime overhead since all offloading decisions are part of the generated binary. However, as Ambit operates on row-wide data, the programmer is required to ensure the operand size to be a multiple or row size. Additionally, all dirty cache lines from the operand rows need to be flushed prior to Ambit operations. Similarly, cache lines from destination rows need to be invalidated. Both of these operations generate additional coherence traffic, which reduces the efficiency of this approach.

*4.3.2 Compiler-Based Approaches*

The main advantage of using compilers for automatic offloading over code annotation is the minimization of programmer burden. This is because the compiler can automatically identify offloading workloads [92], optimize data-reuse [71], and select the PE to execute the offloading workloads [35], [36]. Offloading using compilers also has the potential to outperform manual annotation by better selecting offloading workloads [30]. Like code annotation, this is an offline mechanism and avoids hardware and runtime overhead. However, similar to code annotation, this results in a fixed policy of offloads suited in particular to simple fixed-function memory accelerators which are relatively unconstrained by power and thermal limits [39], [92]. While compilers can be used to predict the performance of an offloading candidate on both host and memory processors, in the case of more complex PIM and NMP designs, compiler-based offloading cannot adapt to changing bandwidth availability, memory access locality, and DRAM temperature, leading to sub-optimal performance, while risking the violation of power and thermal constraints.

Compiler-based techniques have been used with both fixed-function [39], [92], [96] and programmable [27], [40], [70] memory processors. For fixed-function accelerators like vector processors [92] and stencil processors [39], the compiler's primary job is to identify instructions that can be offloaded to the memory PE and configure the operations (e.g., vector size) to match the PE's architecture. When the type of instruction is insufficient to ascertain the benefit of offloading, compiler techniques like CAIRO [42] use analytical methods to quantitatively determine the benefit of offloading by profiling the bandwidth and cache characteristics of instructions offline, at the cost of adding design (compile) time overhead.

Another advantage of using compilers is the ability to efficiently utilize NMP hardware by exploiting parallelism in memory and PEs. For example, [30] achieves 71% better floating-point utilization than hand-written assembly code using loop and memory access optimizations for some kernels. While CAIRO relies on offline profilers and [30] requires some OpenMP directives for annotation, the compiler in PRIMO [92] is designed to eliminate any reliance on third-party profilers or code annotation. Its goal is to reduce communication within multiple memory processors by the efficient scheduling of vector instructions. It can be noticed that when the memory processor is capable of a wider range of operations, compiler-based techniques have to account for aspects like bandwidth saving, cache-locality, and comparative benefit analysis with the host processor [42] while still working offline, which is not easy. Further complications arise when multiple general-purpose processor options exist near memory as runtime conditions become a significant factor in exploiting the benefits of memory-side multi-core processing [27].

*Case Study 1: CAIRO*

CAIRO uses an HMC-based NMP architecture to accelerate graph processing using bandwidth- and locality-aware offloading of HMC 2.0's atomic instructions using a profiler-based compiler. To enable high-bandwidth and low energy, HMC stacks DRAM dies and a CMOS logic die and connects the dies using through-silicon vias (TSVs). Starting with HMC 2.0 specification, HMC has supported 18 atomic computation instructions (shown in Figure 10) on the HMC's logic layer. HMC enables a high level of memory parallelism by using a multiple vault and bank structure. Several properties of graph workloads make HMC-based NMP a perfect choice to accelerate execution. For example, graph computation involves frequent use of read-modify-write operations which can be mapped to HMC 2.0 supported instructions. When performed in HMC, the computation exploits the higher memory bandwidth and parallelism supported by HMC's architecture.

| Type | Data Size | Operation | Return |
|------|-----------|-----------|--------|
| Arithmetic | 8/16-byte | single/dual signed add | w/ or w/o |
| Bitwise | 8/16-byte | swap, bit write | w/ or w/o |
| Boolean | 16-byte | AND/NAND | w/o |
|  |  | OR/NOR/XOR | w/o |
| Comparison | 8/16-byte | CAS-if equal/zero | w/ |
|  |  | greater/less | w/ |
|  |  | compare if equal w/ |  |

**Figure 10.** HMC-atomic instructions in HMC 2.0 [81].

Identifying offloading candidates (instructions that can be offloaded) and choosing suitable memory PEs are important compiler functions for accelerating computation. While this is the main goal with CAIRO, the authors note that additional benefit can be realized when effective cache size and bandwidth saving is considered for offloading decisions. By default, all atomic instructions are considered as offloading candidates and all data for these instructions are marked uncacheable using a cache-bypassing policy from [32]. This technique provides a simple cache coherence mechanism by never caching data that offloading candidates can modify. The underlying idea is that offloading an atomic instruction will inherently save link bandwidth when cache-bypassing is active for all atomic instructions regardless of host or memory-side execution. This is because offloading the instruction to HMC uses fewer flits than executing the instruction on the host CPU which involves the transfer of operands over the off-chip link. Moreover, if the offloaded segment does not have data access locality, it avoids cache pollution, making more blocks available for cache-friendly data. Finally, the NMP architecture allows faster execution of atomic instructions, eliminating long stalls on the host processor.

CAIRO's first step is to identify offloading candidates that can be treated as HMC atomic instructions. Given these instructions, CAIRO then attempts to determine how much speedup can be obtained from offloading these instructions. One of the major factors in determining the speedup is the main memory bandwidth savings from offloading. Since offloading instructions frees up memory bandwidth, it is important to understand how applications can benefit from the higher available bandwidth to estimate application speedup. For applications that are limited by low memory bandwidth, increasing the available memory bandwidth leads to performance improvements (bandwidth-sensitive applications) since these applications exploit memory-level parallelism (MLP). On the other hand, compute-intensive applications are typically bandwidth-insensitive and benefit little from the increase in available memory bandwidth. To categorize their applications, the authors analyze the speedup of their applications after doubling the available bandwidth and conclude that CPU workloads and GPU workloads naturally divide into bandwidth-insensitive and bandwidth-sensitive, respectively. The authors do note that for exceptional cases, CAIRO would need the programmer/vender to specify the application's bandwidth sensitivity.

Given the application's bandwidth sensitivity, CAIRO uses two different analytical models to estimate the potential speedup of offloading instructions. For bandwidth-insensitive applications, they model a linear relationship between the candidate instruction's miss ratio (MR), density of host atomic instructions per memory region of the application ($\rho_{HA}$), and speedup due to offloading the candidate instruction ($SU_{tot}$) as shown in Figure 11(a). In other words, the decision to offload an atomic instruction depends on how often it misses in cache, and how much cache is believed to be available to it. The model incorporates several machine-dependent constants, $C_1$, $C_2$, and $C_3$, shown in Figure 11(b) that are determined using offline micro-benchmarking. When there is positive speedup ($SU_{tot} > 0$), CAIRO's compiler heuristic offloads the instruction to the HMC.

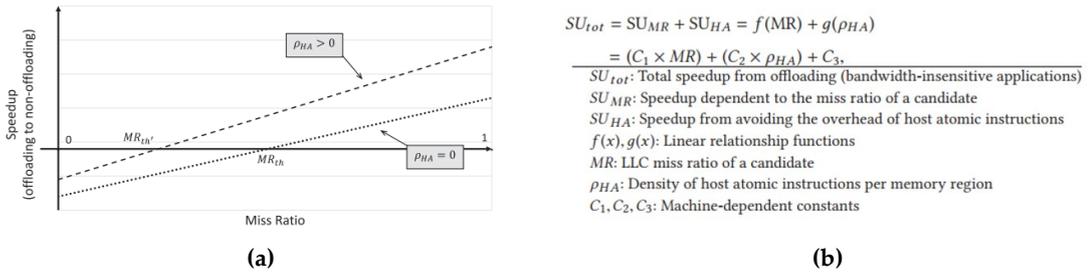

(a)      (b)

**Figure 11.** CAIRO's performance estimation for bandwidth insensitive applications. (a) Speedup given the host atomic instructions per memory region ($\rho_{HA}$) and its miss-ratio. (b) Equation for total speedup, $SU_{tot}$, is the sum of speedup due to avoiding inefficient cache use $SU_{MR}$, and speedup due to avoiding the overhead of executing atomic instructions on the host processor $SU_{HA}$ [42].

On the other hand, bandwidth-sensitive applications can also experience additional speedup by utilizing the extra memory bandwidth due to offloading atomic instructions. Therefore, CAIRO considers bandwidth-savings due to offloading in addition to the miss ratio and density of atomic instructions. As shown in Figure 12, for miss-ratios higher than $MissRatio_H$, the instruction will be offloaded. Similarly, for miss-ratios lower than $MissRatio_L$, the instruction will not be offloaded. However, in the region between $MissRatio_H$ and $MissRatio_L$, additional bandwidth saving analysis is performed before taking the offloading decision. The speedup due to bandwidth saving, $SU_{BW}$ is calculated by estimating the bandwidth savings achieved, $BW'_{saving}$ (Equations 1 and 2). $BW_{reg}$ and $BW_{offload}$ are estimated using hand-tuned equations based on last level cache (LLC) hit ratios, the packet size of offloading an instruction, and the number of instructions (more details in [42]). Constants $M_1$ and $M_2$, are machine-specific and obtained using micro-benchmarking. Similar to the bandwidth-insensitive case, CAIRO's compiler heuristic offloads the instruction when the calculated speedup is greater than 0.

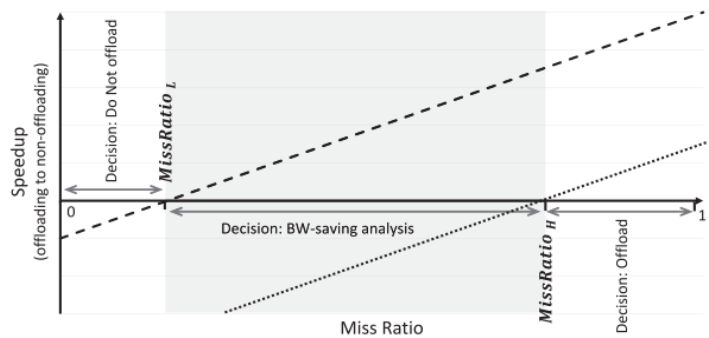

**Figure 12.** CAIRO's performance estimation for bandwidth insensitive applications: For miss-ratios between $MissRatio_L$ and $MissRatio_H$, additional bandwidth-saving analysis is performed before taking the decision to offload [42].

$$SU_{BW} = (M_1 \times BW'_{saving}) + M_2,$$

$SU_{BW}$: Speedup from bandwidth savings (bandwidth-sensitive applications)
$BW'_{saving}$: Normalized bandwidth savings
$M_1, M_2$: Machine-dependent constants

(1)

$$BW'_{saving} = (BW_{reg} - BW_{offload})/BW_{reg} = BW_{saving}/BW_{reg},$$

$BW'_{saving}$: Normalized bandwidth savings
$BW_{reg}$: Regular bandwidth usage (without offloading)
$BW_{offload}$: Bandwidth usage with offloading

(2)

CAIRO's compiler heuristic allows it to consider the most important factors in offloading decisions while avoiding runtime overhead. For bandwidth-insensitive applications, the amount of speedup achieved by CAIRO increases with the cache miss-ratio of the application. For a miss-ratio of greater than 80%, CAIRO doubles the performance of bandwidth-insensitive CPU workloads compared to host execution. For bandwidth-sensitive applications, the amount of speedup increases with the amount of bandwidth saved by CAIRO. For bandwidth savings of more than 50%, CAIRO doubles the performance of GPU workloads. In addition, CAIRO derives energy efficiency from the reduction in data communication inherent to the HMC design. Similar to prior work [28], [32], it extends the HMC instructions and ALU to support floating-point operations without violating power constraints. Despite CAIRO's performance and energy improvements, it has its drawbacks. The miss-ratio and bandwidth saving analysis is machine- and application-dependent and involves considerable design time overhead. Moreover, since the heuristic does not work online, offloading decisions are based on analytical assumptions about runtime conditions. For instance, due to its offline design, CAIRO does not consider the temperature of HMC which can significantly impact the performance [29].

*Case Study 2: A Compiler for Automatic Selection of Suitable Processing-in-Memory Instructions (PRIMO)*

NMP designs with a single PE like CAIRO can focus on offloading the most suitable sections of code to reduce off-chip data movement and accelerate execution. However, when the NMP design uses multiple PEs near memory, designers have to consider the location of data with respect to the PE executing the offloaded workload. The compiler PRIMO [92] is designed to manage such a system. It uses a host CPU connected to an 8-layer HMC unit with 32 vaults. Each of the 32 vaults has a reconfigurable vector unit (RVU) on the logic layer as a PE. The RVU uses fixed-function FUs to execute native HMC instructions extended with an AVX-based ISA, while avoiding the area overhead of programmable PEs. Each RVU has wide vector registers available to facilitate vector operations. They allow a flexible vector width with scalar operands of as little as 4 bytes and vector operands up to 256 bytes. Moreover, multiple RVUs can coordinate to operate on vector lengths of as high as 4096 bytes.

With this architecture, PRIMO has to not only identify offloading candidate code segments to convert to special NMP instructions, but also which execution unit minimizes data movement within memory by exploiting internal data locality. Another way the compiler improves performance is the utilization of the vast number of functional units (FUs) by optimizing the vector length of offloaded instructions for the available hardware. For example, if the RVU architecture can execute instructions up to a vector size of 256 bytes, the compiler will automatically compile offloading candidate code sections with vectors larger than 256 bytes into NMP instructions, whereas operations involving smaller vector sizes are executed by the host CPU. In short, the compiler performs four functions: 1) identify operations with large vector sizes, 2) set these operations' vector size to a supported RVU vector size, 3) check for data dependencies with previously executed instructions, and map instructions to the same vector unit as the previous instructions, and 4) create the binary code for execution on the NMP system. Figure 13(a) shows the code for a *vec-sum* kernel. Figure 13(b) and (c) compare the compiled version of the kernel for an x86 AVX-512 capable processor and an NMP architecture, respectively. The AVX-512 version performs four loads, followed by four more loads and four adds with a vector length of 64 bytes. Finally, four store operations complete an iteration of the loop. In comparison, the NMP code uses just two instructions to load the entire operand data in 256-byte registers, followed by a single 256 bytes add and store operation. The NMP instruction *PIM_256B_LOAD_DWORD V_0_2048b_0 [addr]* loads 256 bytes of data starting at *addr* into register 0 of vault 0. The actual choice of the vault and RVU is made after checking for data dependencies between instructions. The use of large vector sizes allows PRIMO to exploit greater memory parallelism while the locality-aware PE minimizes data movements between vaults.

```
                                mov rax, -16384
                                .LBB0_1:
                                vmovdqu32 zmm8, [rax+c+16576]
                                vmovdqu32 zmm4, [rax+c+16512]
                                vmovdqu32 zmm3, [rax+c+16448]
                                vmovdqu32 zmm0, [rax+c+16384]         mov rax, -16384
                                vpaddd zmm9, zmm0, [rax+b+16384]      .LBB0_1:
for(int i=0; i<N; i++)          vpaddd zmm6, zmm3, [rax+b+16448]      PIM_256B_LOAD_DWORD V_0_R2048b_0, [rax+b+16384]
    c[i] = a[i] + b[i];         vpaddd zmm3, zmm4, [rax+b+16512]      PIM_256B_LOAD_DWORD V_0_R2048b_1, [rax+c+16384]
                                vpaddd zmm0, zmm8, [rax+c+16576]      PIM_256B_VADD_DWORD V_0_R2048b_1, V_0_R2048b_0, V_0_R2048b_1
     (a) C Code                 vmovdqu32 [rax+a+16384], zmm0         PIM_256B_STORE_DWORD [rax+a+16384], V_0_R2048b_1
                                vmovdqu32 [rax+a+16448], zmm3         add rax, 4096
                                vmovdqu32 [rax+a+16512], zmm6         jne .LBB0_1
                                vmovdqu32 [rax+a+16576], zmm9
                                add rax, 4096                              (c) NMP ASM Code
                                jne .LBB0_1

                                   (b) AVX-512 ASM Code
```

**Figure 13.** PRIMO Code Generation [92].

While identifying vector instructions is common to all compiler-based management techniques, PRIMO achieves better FU utilization and minimizes internal data movement by optimizing instruction vector size and selecting PEs considering the location and vector size of the operand data. It also manages to completely automate the process and eliminate the need to annotate sections of code or run offline profilers. In addition, PRIMO performs all optimizations at design time which allows it to eliminate runtime overhead. PRIMO achieves an average speedup of 11.8× on a set of benchmarks from the PolyBench Suite [99], performing the best with large operand vector sizes. But such benefits come at the cost of a static policy that is unable to react to dynamic system conditions like available memory bandwidth, variations in power dissipation, and temperature. Additionally, the energy efficiency of specialized hardware and instructions need to be studied further.

*4.3.3 Online Heuristic*

Online heuristic-based policies use human expert knowledge of the system to make offloading decisions at runtime. To this end, additional software or hardware is tasked with monitoring and predicting the future state of the system at runtime, to inform the offloading workload identification, determining the memory PE, and timing control. The online nature of heuristic policies enables the policy to adapt to the runtime state, but it also increases its complexity and runtime overhead. The quality of decisions depends on the heuristic designer's assumptions about the system. When the assumptions are incomplete or incorrect, online heuristics will suffer from unexpected behavior.

Online heuristics are used widely when the goal is to accelerate an entire data-intensive application on general purpose memory processors [27], [38], [71], [72], [74]. In addition, online heuristics may be used in combination with code annotation and compiler-based methods [27], [75]. For example, PEI [28] extends the host ISA with new PIM-enabled instructions to allow programmers to specify possible offloading candidates for PIM execution. These instructions are offloaded to PIM only if a cache hit-counter at runtime suggests inefficient cache use, otherwise the instructions are executed by the host CPU. Similarly, TOM [27] requires the programmer to identify sections of CUDA code as candidate offloading blocks. The final offloading decision involves an online heuristic to determine the benefit of offloading by estimating bandwidth saving, and co-locate data and computation across multiple HMC modules. Different from CAIRO where the bandwidth saving analysis is performed entirely before compilation, TOM uses heuristic analysis of bandwidth saving both before and after compilation. It marks possible offloading candidates during a compiler pass, but the actual offloading decision is subject to runtime monitoring of dynamic system conditions like PE utilization and memory bandwidth utilization. Other methods that rely on online heuristics include [38] which uses a two-tier approach to estimating locality and the energy consumption of offloading decision, [74], which performs locality-aware data allocation and prefetching, and [29], which optimizes for thermally feasible NMP operation by throttling the frequency of issuing offloads and size of CUDA blocks offloaded to an HMC-based accelerator.

*Case Study 1: CoolPIM: Thermal-Aware Source Throttling for Efficient PIM Instruction Offloading*

As noted in Section 4.1.3, 3D-stacked memory is vulnerable to thermal problems due to high power densities and insufficient cooling, especially at the bottom layers. In fact, with a passive heatsink, HMCs cannot operate at their maximum bandwidth even without PIM and NMP (see Figure 14) [29], [100]. A particular problem in these systems is that heating exacerbates charge leakage in DRAM memory cells which demands more frequent refreshes to maintain reliable operation. Therefore, offloading in NMP, among other online runtime conditions, requires awareness of the memory system's temperature. CoolPIM [29] attempts to solve this problem using an online-heuristic based mechanism to control the frequency of offloading under the thermal constraints of an NMP system.

CoolPIM considers a system that has a host GPU executing graph workloads with an HMC memory module capable of executing HMC 2.0 atomic instructions. In order to allow the offloading of atomic operations written in Nvidia's CUDA to HMC, the compiler is tweaked to generate an additional executable version for each CUDA block of computation. This alternative HMC-enabled version has the HMC's version of atomic instructions to be executed near-memory if the heuristic decides to offload it. All atomic instructions in HMC-enabled blocks execute on functional units on the HMC's logic layer as provided by the HMC 2.0 Specification [81]. It must be noted that role of the compiler is not to make offloading decisions, but rather to generate a set of offloading candidates that the online heuristic can select from at runtime.

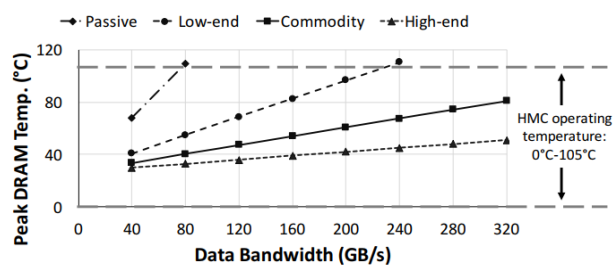

**Figure 14.** Peak DRAM temperature with various data bandwidth and cooling methods (passive, low-end, commodity, and high-end) [29].

The main goal of CoolPIM is to keep the HMC unit under an operational maximum temperature. The HMC module includes a thermal warning in response packets whenever the surface temperature of the module reaches 85°C. Whenever the thermal warning is received, CoolPIM throttles the intensity of offloading by reducing the number of HMC-enabled blocks that execute on the GPU. To this end, CoolPIM introduces a software-based throttling and hardware-based throttling technique illustrated in Figure 15. The software-based throttling technique controls the number of HMC-enabled blocks that execute on the GPU using a thermal warning interrupt from the HMC. It implements a token-based policy where CUDA blocks request a token from the pool before starting execution. If a token is available, the block acquires it and the offloading manager executes the NMP version of the block. However, if a thermal warning is encountered, the token pool is decremented, which effectively decreases the total number of blocks that can be offloaded to the memory PE.

On the other hand, the hardware-based throttling technique controls the number of warps per block that are offloaded to the memory PE as a reaction to the thermal warning interrupt. Unlike the software-based method which requires that all warps in executing thread blocks finish before the throttling decisions take effect, the hardware-based mechanism allows the system to react faster to the temperature warning by changing the number of warps with NMP-enabled instructions during runtime. The hardware-based control is achieved by adding a hardware component to the system which can replace NMP-enabled instructions with their CUDA counterparts during the decoding process. The HMC-disabled warps, with NMP instructions replaced with CUDA instructions will then execute entirely on the host GPU to help manage the rising HMC temperature. Figure 15 details the implementation of the policy.

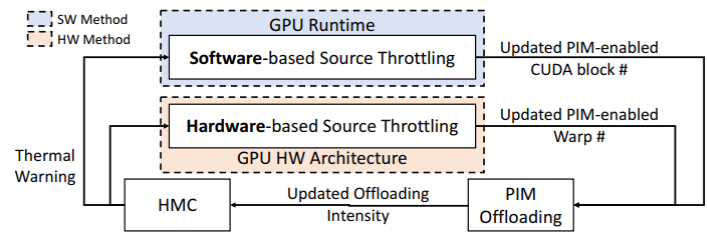

**Figure 15.** Overview of CoolPIM [29].

The use of an online heuristic allows CoolPIM to adapt to system conditions at runtime and control temperature even when an application goes through different phases. The additional system awareness not only improves the offloading decisions to better realize the benefits of NMP but also develops a more proactive approach to management. CoolPIM improves performance by up to 40% compared to both host execution and naïve offloading, while effectively managing memory temperature. While an increase in the heuristic's complexity is expected to produce better offloading decisions and performance improvement, it is necessary that runtime overhead remain manageable. While CoolPIM manages to do that, it does not consider data locality and bandwidth saving while taking offloading decisions. Similarly, it does not consider energy savings except as a by-product of minimizing DRAM refresh rate.

*Case Study 2: Making Better Use of Processing-in-Memory Through Potential-Based Task Offloading*

Although works under the code annotation and compiler categories use profiling and analysis to understand the effects of offloading on key objectives, the generated results may be affected by runtime conditions related to input data, concurrent workloads, and other dynamic policies. Kim et al. [38] look at the number of L2 misses, memory stalls, and power in their online heuristic policy to determine offload decisions to optimize performance, energy, and power. They consider an NMP system with a host GPU and a trimmed-down GPU as the memory PE in the logic layer of an HBM unit. The GPU used as a memory PE has a lower number of streaming multiprocessors (SM) to accommodate the thermal constraints on the logic layer of the HBM. To accomplish their goals, the authors divide the heuristic into two stages: the first stage determines if the decision meets a time-energy constraint ($OC_{T-E}$), and if it passes, the second stage will determine if it passes a time-power constraint ($OC_{T-P}$). Offloading is performed only if the conditions in both stages are met. This process is represented in Figure 16.

The first stage starts by profiling the application for the number of L2 misses per instruction (L2MPI) and the number of cycles L1 was stalled due to a request to L2 (L1S). The reason behind this choice is that the higher the value of each of these metrics, the slower the task will run on the host GPU and the better it will perform on the high memory bandwidth near-memory GPU. The two values are used to determine if the task should be offloaded using a simple linear model derived through empirical data from profiling (Equation 3). If the value of this model exceeds a threshold (TH), then the task is considered as an offloading candidate based on the time-energy constraint.

$$58 \times L2MPI + 0.22 \times L1S + 0.005 > TH \quad (3)$$

If a task passes the first stage, the next stage is to determine the impact of the offloading decision on power consumption. Specifically, power is treated as a constraint with a fixed budget of 55W with a high-end-server active heat sink. The power model for estimating the task's power consumption in NMP execution is based on several assumptions. Idle power is ignored in the work. Static power is scaled by the ratio of the areas of the host and memory GPU. For the dynamic power estimation, different components of the host power are scaled down using information from profiling the application. Equation 4 shows how the relationship is modeled.

$$PIM_{SM} = \alpha \times \left((ActiveSM_{Host} + L2_{Host} + ICNT_{Host}) + 0.25 \times (MC_{Host} + DRAM_{Host})\right) \quad (4)$$

All dynamic power components are scaled by an experimentally determined parameter $\alpha$. These components include the power consumed by active streaming-multiprocessor (SM) ($ActiveSM_{Host}$), the L2 cache in the host processor ($L2_{Host}$), and the interconnect network ($ICNT_{Host}$). The DRAM and memory controller (MC) power components ($MC_{Host}$ and $DRAM_{Host}$) are further scaled down by 0.25, which is the experimentally determined ratio between DRAM and active SM power. The value of $\alpha$ is estimated by profiling the application. The L2MPI and percentage of active SMs to idle SMs is used to infer how much the application would benefit from executing on the memory side PE. The assumption is that large L2MPI values indicate inefficient use of cache, and wastage of memory bandwidth. Similarly, a low percentage of active cores indicates that the application will not experience a slow-down due to the lower number of compute units in the memory PE.

Finally, another linear model based on these assumptions of power scaling is used to check if the use of DVFS will make some offloading decisions feasible within the power constraint. If all tests are passed, the task is offloaded to the memory-side GPU.

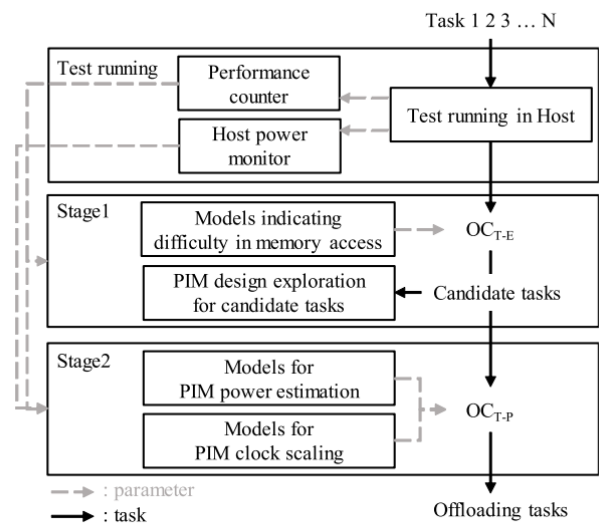

**Figure 16.** Overview of the offloading policy used in [38]. The offloading decision is separated into two stages. The first stage checks whether it would pass a time-energy constraint ($OC_{T-E}$) while the second stage checks whether it passes a time-power constraint ($OC_{T-P}$). If it passes both, the task is offloaded onto the memory module.

This technique comprehensively considers the energy, power, and performance aspect of offloading to power constrained NMP systems. Compared to host-only execution, the proposed technique achieves a 20.5% increase in instructions per cycle (IPC) and a 67.2% decrease in energy delay squared product (ED2P). The approach comes with additional design time and runtime overhead compared to code annotation and compiler-based methods. Additionally, it is based on strict assumptions about the system dynamics derived from profiling characteristic applications. Practically, the different system components may interact in a variety of unpredictable ways, breaking the assumption of stationarity.

## 5. Conclusions and Future Challenges and Opportunities

PIM and NMP architectures have the potential to significantly reduce the memory wall and enable future big data applications. However, it has been demonstrated that naïve use of these DCC systems can run into thermal issues and even reduce the performance of the overall system. This has led many to investigate offloading management techniques that are able to identify low data locality instructions or react to thermal emergencies. In this paper we have organized these works based on the optimization objective, optimization knobs, and the type of technique they utilize. However, there are several challenges and opportunities for resource management of PIM/NMP substrates related to generalizability, multi-objective considerations, reliability, and the application of more intelligent techniques, e.g., machine learning (ML), as discussed below:

- Due to the large variations in DCC architectures proposed to date, the management policies have been very architecture- and application-specific. For example, a policy for near-memory graph accelerator involves the offloading of specific graph atomic operations, a policy for a stencil accelerator involves the offloading of stencil operations, and so on. This could lead to technology fragmentation, lower overall adoption, and inconsistent system improvements. As an example, an NMP system designed to exploit the high density of connections in small graph areas fails to extract significant speedup when graph connections are more uniform [33]. Future DCC architectures and resource management policies need to explore the generalizability of these systems.
- Nearly all past work has focused on the efficient use of PEs across the system, i.e., they offload tasks which minimize data movement between the host processor and memory. However, high frequency of offloading can cause the memory chip to overheat and lead to a complete shutdown. The issue is addressed reactively in [29] but more proactive and holistic resource management approaches are needed to consider both thermal- and performance-related objectives together.
- Reliability has yet to be considered in the management policies for DCC systems [59]. There is no analysis of the impact of reliability concerns in emerging NVM substrates or DRAM cells on the efficacy of PIM/NMP offloading strategies. Due to deep nanometer scaling, DRAM cell charge retention is becoming increasingly variable and unpredictable. Similarly, the use of unproven and new NVM technologies that are susceptible to disturbances during non-volatile programming brings some level of uncertainty at runtime. Resource management techniques need to be designed in a manner that is robust to these reliability issues in memory substrates, when making decisions to offload to memory PEs.
- ML based applications have exploded in recent years. ML's potential has been demonstrated for identifying offloading targets [71] using a simple regression-based model with cache performance metrics as the input. More generally, ML techniques like reinforcement learning has proven successful in improving performance by intelligently scheduling workloads on heterogenous systems [101]. As we adopt more general architectural designs, management policies will need to account for the diversity of applications and variability of processing resources. On the other hand, IoT devices have great potential to use ML for smart operation, but they lack the resources for training ML models on large datasets. Recent work [102] has shown that executing ML algorithms using near-data vector processors in IoT devices can significantly improve performance. Hence, a promising direction is to leverage the DCC approach to empower IoT devices to process data locally to improve privacy and reduce latency.
- Heterogenous manycore architectures running multi-threaded applications result in complex task mapping, load balancing, and parallelization problems due to the different PEs. Recently, complex network theory, originally inspired by social networks, has been successfully applied to analyze instruction and data dependency graphs and identify "clusters" of tasks to optimally map instructions to PEs [103]. Similarly, complex network theory can be extended to include the PIM and NMP domain in order to optimize software, data orchestration, and hardware platform simultaneously.

**Author Contributions:** Conceptualization, K.K., S.P., and R.K.; investigation, K.K.; writing—original draft preparation, K.K and R.K.; writing—review and editing, K.K., S.P., and R.K; visualization, K.K.; supervision, S.P. and R.K.; project administration, S.P. and R.K. All authors have read and agreed to the published version of the manuscript.

**Funding:** This research was supported by the National Science Foundation (NSF) under grant number CCF-1813370.

**Conflicts of Interest:** The authors declare no conflict of interest.